\documentclass[english,preprint, aps, prx, notitlepage]{revtex4-1}
\usepackage[T1]{fontenc}
\usepackage[latin9]{inputenc}
\usepackage{geometry}
\usepackage{xcolor}
\usepackage{babel}
\usepackage{array}
\usepackage{booktabs}
\usepackage{textcomp}
\usepackage{amsmath}
\usepackage{graphicx}
\PassOptionsToPackage{normalem}{ulem}
\usepackage{ulem}
\usepackage[unicode=true,
 bookmarks=true,bookmarksnumbered=false,bookmarksopen=false,
 breaklinks=true,pdfborder={0 0 0},pdfborderstyle={},backref=false,colorlinks=false]
 {hyperref}
\hypersetup{pdftitle={Efficient molecular density functional theory using generalized spherical harmonics expansions},
 pdfauthor={Lu Ding, Maximilien Levesque, Daniel Borgis and Luc Belloni}}
\usepackage{breakurl}

\makeatletter

\providecommand{\tabularnewline}{\\}
\providecolor{lyxadded}{rgb}{0,0,1}
\providecolor{lyxdeleted}{rgb}{1,0,0}
\usepackage[T1]{fontenc}

\makeatother

\begin{document}

\title{High-throughput free energies and water maps for drug discovery by
molecular density functional theory}

\author{Sohvi Luukkonen}

\affiliation{Maison de la Simulation, USR 3441 CNRS-CEA-Université Paris-Saclay,
91191 Gif-sur-Yvette, France}

\author{Luc Belloni}

\affiliation{LIONS, NIMBE, CEA, CNRS, Université Paris-Saclay, 91191 Gif-sur-Yvette,
France}

\author{Daniel Borgis}

\affiliation{Maison de la Simulation, USR 3441 CNRS-CEA-Université Paris-Saclay,
91191 Gif-sur-Yvette, France}

\affiliation{PASTEUR, Département de chimie, École normale supérieure, PSL University,
Sorbonne Université, CNRS, 75005 Paris, France}

\author{Maximilien Levesque}

\affiliation{PASTEUR, Département de chimie, École normale supérieure, PSL University,
Sorbonne Université, CNRS, 75005 Paris, France}
\email{maximilien.levesque@ens.fr}

\date{\today}
\begin{abstract}

The hydration or binding free energy of a drug-like molecule is a
key data for early stage drug discovery. Hundreds of thousands of
evaluations are needed, which rules out the exhaustive use of atomistic
simulations and free energy methods. Instead, the current docking
and screening processes are today relying on numerically efficient
scoring functions that lose much of the atomic scale information and
hence remain error-prone. In this article, we show how a probabilistic
description of molecular liquids as implemented in the molecular density
functional theory predicts hydration free energies of a state-of-the-art
benchmark of small drug-like molecules within 0.5 kJ/mol (0.1 kcal/mol)
of atomistic simulations, along with water and polarization maps,
for a computation time compatible with screening and docking.
\end{abstract}
\maketitle

\section{Introduction}

The development of a new drug is a long and expensive process, consuming
in average 10 to 12 years and between 2.5 and 3.0 billion dollars~\cite{dimasi_2016}.
Structure-based drug design process starts with the identification
of a protein related to the disease, called the target, whose activity
needs to be modulated with the drug to be found. Once the target is
found, the early stages of drug discovery consist in finding few potential
drug leads via \textit{\emph{back and forth between }}\textit{in silico}
methods that start from databases of millions of small drug-like molecules.
The screening, i.e., ranking, of drug candidates is then done by a
score, mimicking the binding free energy between the small ligand
molecule and the protein when the ligand is docked in the active site
of the protein.

However, computing free energies of a process such as binding or solvation
is difficult as it requires the sampling of all possible states that
could be visited during the transformation. Nevertheless, free energy
methods that use alchemical transformations exist. They range from
simple exponential averaging (EXP) introduced by Zwanzig~\cite{zwanzig_hightemperature_1954}
60 years ago to more sophisticated method employing non-physical intermediate
states such as the thermodynamic integration (TI)~\cite{kirkwood_statistical_1935},
free energy superposition, the Bennett acceptance ratio (BAR)~\cite{bennett_efficient_1976},
the weighted histogram analysis method (WHAM)~\cite{kumar_1992}
or the multistate Bennett acceptance ratio (MBAR)~\cite{shirts_2008}.
All these methods require multiple ergodic molecular dynamics (MD)
or Monte-Carlo (MC) simulations for a single free energy estimate.
In other words, they require multiple simulations to be run, typically
few tens, eventually in parallel, with the associated few tens of
simulation time. These methods and the protocol associated for producing
the free energy estimate defines today's standard in terms of free
energy predictions.

In the case of drug discovery, binding free energies need to be calculated
in water as the protein-ligand interaction often takes place in an
aqueous environment. The direct calculation of binding free energies
in the solvent is thus especially difficult because all the configurations
of the solvent molecules have to be sampled in addition to protein
and ligand configurations. However the binding free energy between
two molecules A and B in solution can be rigorously expressed in terms
of the solvation free energies of the complex AB, the molecule A,
the molecule B and of the binding free energy in vacuum :
\begin{eqnarray}
\Delta G_{\textrm{bind,solv}}^{\textrm{AB}} & = & \Delta G_{\textrm{solv}}^{\textrm{AB}}-(\Delta G_{\textrm{solv}}^{\textrm{A}}+\Delta G_{\textrm{solv}}^{\textrm{B}})+\Delta G_{\textrm{bind,vac}}^{\textrm{AB}}.\label{eq:bindingfreeenergy}
\end{eqnarray}
Since the binding in vacuum, $\Delta G_{\textrm{bind,vac}}^{\textrm{AB}}$,
is much easier to compute, the difficulty in obtaining a binding free
energy in solution is moved to the solvation free energy calculation
(SFE), that is, in $\Delta G_{\textrm{solv}}$ of equation~\ref{eq:bindingfreeenergy}.
SFE calculations can be done with explicit or implicit models of water.
The theory behind alchemical transformations is exact, thus the only
approximations are the choice of a force field for describing the
interaction the solute-solvent interaction and numerical errors. However,
in explicit solvent alchemical transformations are computationally
expensive as multiple ergodic simulations are needed to accumulate
enough uncorrelated data points to have an acceptable statistical
error. Obtaining a single solvation free energy of a small drug-like
molecule requires hundreds cpu.h with an explicit solvent method,
which is too slow for the screening of millions of drug-like ligands
of the early stage \textit{in silico} drug discovery process. Hence
they can only be used in ``late-stage'' \textit{in silico} drug
design. Therefore, one uses parameterized implicit solvent models
or scoring functions for the initial stages of the screening process.
Implicit solvent alchemical transformations estimate SFEs in few tens
of seconds at most. Scoring functions are fast, more or less empirical
functions that use some molecular descriptors of the ligands to rank
them~\cite{kitchen_2004}. These highly parameterized functions are
fast enough for processing large databases, requiring few minutes
at most per score~\cite{zhou_2007}. In some cases, they work really
well, but the lack of physical support and insight was shown by recent
reviews to lead to unpredictable qualities~\cite{chaput_2017,chaput_2016}.

Alternative so called ``end-point'' methods exist, ie. methods without
alchemical intermediates, such as Watermap~\cite{abel_2008,young_2007}
where the water-density is obtained from an explicit solvent MD simulation
and is then injected into a functional that estimates the binding
free energy. Some other methods are based solely on electrostatics
such as AquaSol~\cite{koehl_2010}, which uses a dipolar Poisson\textendash Boltzmann\textendash Langevin
formalism to calculate the solvent density and solvation free energy. 

Other ``end-point'' approaches have roots in the liquid state theories.
They are based on the Ornstein-Zernike equation (OZ) solved within
the integral equation or classical density function theory (cDFT)
formalism. Among them, reference interaction site model (RISM) and
3D-RISM~\cite{chandler_optimized_1972,hirata-rossky81,Beglov-Roux97,kovalenko-hirata98,kovalenko_hydration_2000,kovalenko_potential_1999}
are simplified versions which ignore the full molecular description
of the solvent and replace it by site-site correlations only. They
enable the calculation of the site densities, such as the water-oxygen
or hydrogen densities and of the solvation free energy by minimizing
a site-functional~\cite{liu_site_2013} or solving the 3D-RISM equations.
These methods are having large success and are gaining momentum because
of their good balance between precision, simplicity and speed. Nevertheless,
they rely on simplifying molecules to a set of sites correlated together,
not on a full molecular description and hence they use multiple compromises
with purely phenomenological corrections. 

On the other hand, the molecular density functional theory (MDFT)
\cite{ramirez05-CP,gendre_classical_2009,chandler_density_1986,chandler_density_1986-1,ramirez_density_2002,zhao_molecular_2011}
and molecular integral equation theory (MIET) are the only methods
based on liquid state theories that keep the full molecular picture
by solving the molecular Ornstein-Zernike equation (MOZ). It is diagrammatically
consistent. Since Ding et al.~\cite{ding_2017}, MDFT can be solved
efficiently and rigorously at the hypernetted chain (HNC) level of
approximation. In this paper, we assess of MDFT in the HNC approximation
(MDFT-HNC) to predict solvation free energies of small drug-like molecules.
MDFT-HNC is a rigorous, lowest level of the hierarchy of functionals
for the MDFT: It can thus be used as a starting point to be systematically
improved. While this can be done by developing so-called bridge functionals~\cite{levesque12_1,jeanmairet_molecular_2015,cageat_2018,zhao-wu11,zhao-wu11-correction,liu_site_2013},
this paper is dedicated to MDFT-HNC in its ``crudest'' approximation
and its capacity to predict solvation free energies of small drug-like
molecules. It lays the foundations of the future of MDFT for the early
stage \textit{in silico} drug discovery.

In section~II, we describe the MDFT framework. In section~III, we
present reference calculations using state-of-the-art alchemical transformations.
In section IV, we compare the references to MDFT-HNC for what concerns
solvation free energies and solvation structure of drug-like molecules.
Conclusions and perspectives are drawn in section~V.

\section{Molecular Density Functional Theory}

The molecular density functional theory of classical, molecular fluids,
computes rigorously and efficiently the solvation free energy and
equilibrium solvent density around a solute. MDFT is a cousin of the
well-known Kohn-Sham density functional theory for electrons, extended
to finite temperature in the grand canonical ensemble by Mermin~\cite{hohenberg_inhomogeneous_1964,kohn_self-consistent_1965,mermin_thermal_1965}.
In the classical density functional theory formalism, the solvation
free energy $\Delta G_{\textrm{solv}}$ (SFE) is defined as the difference
between the grand potential $\Omega$ of the solvated system, with
the grand potential $\Omega_{b}$ of the bulk solvent:

\begin{equation}
\Delta G_{\textrm{{solv}}}=\Omega-\Omega_{b}=\textrm{min}\left\{ \mathcal{F}[\rho]\right\} =\mathcal{F}[\rho_{\text{eq}}],\label{eq:minimisation}
\end{equation}
where $\mathcal{F}[\rho]$ is the functional to be minimized, $\rho=\rho(\boldsymbol{r},\omega)$
the molecular solvent density function with $\boldsymbol{r}$ a three
dimensional vector and $\omega$ the Euler angles $(\theta,\phi,\psi)$,
characterizing the position and the orientation of the rigid solvent
molecule, typically the TIP3P model for water \cite{TIP3P} and $\rho_{eq}$
is the equilibrium solvent density.~

The MDFT functional $\mathcal{F}$, to be minimized, is made of three
parts: 

\begin{equation}
\mathcal{F}=\mathcal{F}_{{\rm id}}+\mathcal{F}_{{\rm ext}}+\mathcal{F}_{{\rm exc}},\label{eq:Fid+Fext+Fexc}
\end{equation}
where $\mathcal{F}_{{\rm id}}$ is the ideal term of a fluid of non-interacting
particles, $\mathcal{F}_{\textrm{ext}}$ is the external term induced
by the solute (the protein, ligand or complex embedded in water),
and $\mathcal{F}_{{\rm exc}}$ is the excess term that includes structural
correlations between solvent molecules. The ideal term, coming solely
from the entropy of mixing of the solvent molecules, reads

\begin{equation}
\mathcal{F}_{{\rm id}}=k_{{\rm B}}T\int\mathrm{d}\mathbf{r}\mathrm{d}\mathbf{\omega}\left[\rho(\mathbf{r},\mathbf{\omega})\ln\left(\dfrac{\rho(\mathbf{r},\mathbf{\omega})}{\rho_{{\rm bulk}}}\right)-\Delta\rho(\mathbf{r},\mathbf{\omega})\right],\label{eq:1.1}
\end{equation}
where $k_{{\rm B}}T$ is the thermal energy, $\textrm{d\ensuremath{\boldsymbol{r}}}\equiv\textrm{dxdydz}$,
$\textrm{d\ensuremath{\omega\equiv}d\ensuremath{\cos\theta}d\ensuremath{\phi}d\ensuremath{\psi}}$
and $\Delta\rho(\mathbf{r},\mathbf{\omega})\equiv\rho(\mathbf{r},\mathbf{\omega})-\rho_{{\rm bulk}}$
is the excess density over the bulk homogeneous density $\rho_{\textrm{bulk}}\equiv n_{\textrm{bulk}}/8\pi\text{\texttwosuperior}$.
$n_{\textrm{bulk}}$ is typically 0.033~molecule per \AA$^{3}$
for water at room conditions ($\equiv1$~kg/L). $8\pi^{2}$ is the
angular normalization constant. The external contribution comes from
the interaction potential $v_{{\rm ext}}$ between the solute molecule
and a solvent molecule. It reads 
\begin{equation}
\mathcal{F}_{{\rm ext}}=\int\mathrm{d}\mathbf{r}\mathrm{d}\mathbf{\omega}\rho\left(\mathbf{r},\mathbf{\omega}\right)v_{{\rm ext}}\left(\mathbf{r},\mathbf{\omega}\right),
\end{equation}
where $v_{\textrm{ext}}$ is the interaction energy between a solute
and a solvent molecule that is made of a van-der-Waals term, typically
Lennard-Jones, and electrostatic interactions. Those are the same
non-bonded force field parameters as in a molecular dynamics simulation.
For now and in what follows, the MDFT does not use the intramolecular
force field parameters as the solute and the solvent are considered
rigid for the sake of the simplicity of the demonstration. For the
solvent, we use the TIP3P \cite{TIP3P} water model as it is the most
commonly found in drug design applications and thus in reference simulations
detailed below.~The final, excess term describes the effective solvent-solvent
interactions, may be written as a density expansion around the homogeneous
bulk density $\rho_{\textrm{bulk}}$:

\begin{align}
\mathcal{F}_{{\rm exc}} & =-\frac{k_{\text{B}}T}{2}\int d\mathbf{r}_{1}\mathrm{d}\mathbf{\omega}_{1}\int d\mathbf{r}_{2}\mathrm{d}\mathbf{\omega}_{2}\Delta\rho\left(\mathbf{r}_{1},\mathbf{\omega}_{1}\right)c^{(2)}\left(r_{12},\mathbf{\omega}_{1},\mathbf{\omega}_{2}\right)\Delta\rho_{2}\left(\mathbf{r}_{2},\mathbf{\omega}_{2}\right)+\mathcal{F}_{{\rm b}}\nonumber \\
 & =-\frac{k_{\text{B}}T}{2}\int d\mathbf{r}_{1}\mathrm{d}\mathbf{\omega}_{1}\Delta\rho\left(\mathbf{r}_{1},\mathbf{\omega}_{1}\right)\gamma\left(\mathbf{r}_{1},\mathbf{\omega}_{1}\right)+\mathcal{F}_{\textrm{b}}\nonumber \\
 & =\mathcal{F}_{{\rm HNC}}+\mathcal{F}_{{\rm b}},
\end{align}
where $c^{(2)}\left(\boldsymbol{r}_{12},\mathbf{\omega}_{1},\mathbf{\omega}_{2}\right)$
is the homogeneous solvent-solvent molecular direct correlation function,
$\mathcal{F}_{\textrm{b}}$ the bridge functional and $\gamma\equiv c^{(2)}*\Delta\rho$
is the indirect solute-solvent correlation defined as the spatial
and angular convolution of the excess density with the direct correlation
function. If one cuts the expansion to order two in excess density,
that is, if one cancels the bridge functional~\cite{vanleeuwen_1959},
one finds that the MDFT functional produces at its variational minimum
the well-known HNC relation for the solute-solvent distribution function:

\begin{equation}
g(\boldsymbol{r},\omega)=\frac{\rho_{\text{eq}}(\boldsymbol{r},\omega)}{\rho_{\textrm{bulk}}}=e^{-\beta v_{\textrm{ext}}(\boldsymbol{r},\omega)+\gamma},
\end{equation}
where $\beta\equiv\frac{1}{k_{B}T}$. Therefore, we call the first
term of the excess functional the HNC functional. The function $c^{\text{(2)}}$
of the bulk solvent for a given temperature and pressure is an input
in the present theory and is provided by previous Monte Carlo simulations
coupled to integral equations calculations \cite{puibasset_bridge_2012,belloni_2017}
carefully corrected for finite-size effects~\cite{belloni2017finite}.
Higher-order correlation functions could be computed but usually are
not numerically feasible. The rest of the excess term, the so-called
bridge functional can be approximated rigorously or empirically~\cite{levesque12_1,jeanmairet_molecular_2015,cageat_2018}.
However in this paper we benchmark MDFT as a prospective tool for
drug design applications, at its lowest level of accuracy : the MDFT-HNC,
\textit{i.e.} without a bridge functional. This HNC level can only
be improved by adding bridge functionals. Contrary to other liquid
state theories, the MDFT-HNC is a rigorous basis one can only improve.

This theory and corresponding algorithms are implemented into a in-house
high performance code that predicts a hydration free energy in few
seconds to minutes. It depends on the simulation cell size and spacial
and angular resolutions~\cite{ding_2017}. The MDFT code includes,
as expected, partial molar volume correction (also called PC or ISc
correction) to the solvation free energy~\cite{sergiievskyi_solvation_2015,sergiievskyi_fast_2014}.

\section{Reference calculations from simulation}

To assess the quality of the hydration free energies (HFE) predicated
in few minutes by MDFT-HNC, we use the FreeSolv database~\cite{mobley_2013,matos_2017_freesolv}.
FreeSolv concatenates hydration free energies of 642 neutral drug-like
molecules, as measured experimentally and calculated with state-of-the-art
MBAR from explicit solvent molecular dynamics simulations with the
Generalized Amber force field (GAFF)~\cite{wang_gaff_2004} version
1.7 with AM1-BCC~partial charges \cite{jakalian_2000,jakalian_2002}
and TIP3P model~\cite{TIP3P} of water. In these simulations, the
solute is flexible, that is, its internal degrees of freedom are free.
How the hydration free energy depends on the ensemble of conformers
for a given molecule is a statistical mechanical problem in itself,
already tackled for instance through the exploration of so-called
end-point calculations. The illustration of this is given in fig.~\ref{fig:correlations},
where one sees HFEs of all Freesolv molecules as calculated with a
flexible solute, i.e., with a mixture of solute conformers, and as
calculated for a rigid solute, that is, for a single solute conformer.
We used the initial conformations given in FreeSolv database. As summarized
in Table~\ref{tab:corr_summary}, the mean unsigned error with respect
to experimental changes from 4.3~kJ/mol to 5.4~kJ/mol because of
the single point calculation and the Pearson R coefficient, that is
the correlation, change from 0.94 to 0.89. It is not the point of
this manuscript to review those methods to take into account the flexibility
or mix of conformers. For the sake of the simplicity of the demonstration,
we performed rigid solute (a single conformer) insertion/destruction
Monte Carlo (MC) simulations of all molecules in the databases coupled
with the Bennett acceptance ration (BAR) analysis \cite{belloni2017finite,belloni_2017,bennett_efficient_1976,belloni_2014}.
Simulations details can be found in the supplementary information
(SI). Since we use the same force fields and conformers in the simulations
and in MDFT, i.e., since we lie in the same force field approximation,
the fully atomistic simulation based alchemical transformation calculations
can be considered as reference calculations. The computation cost
of such a reference state-of-the-art simulation is of the order of
one day or few hours with the most recent GPU compatible codes.

Additionally, we performed implicit solvent calculations with Yank
(an open source Replica Exchange MC code for alchemical transformation
\cite{eastman_2010,eastman_2010-1,eastman_2013,friedrichs_2009,shirts_2007,shirts_2008,chodera_2011,yank})
on the whole FreeSolv database. In the implicit solvent calculations
the rigid solute is also described with GAFF, while the implicit solvent
is of type GB-OBC(II)~\cite{onufriev_2004}. The simulation details
are given in SI. Knight and Brooks have published a review of implicit
solvent models for a previous version of the FreeSolv database \cite{knight_2011}.
Our calculations are much more limited than Knight and Brooks as we
only seek to illustrate typical results, but our conclusions are in
full agreement with theirs.

We also evaluated the capacity of MDFT to predict the equilibrium
solvent structures around a solute, the so-called \emph{water maps}.
To this end, we performed reference simulations of a rigid solute
in water chosen to be representative of the molecules of the FreeSolv
database : the quinoline (C9H7N). It has the correct average size,
contains a hetero-atom and a cycle. 

\section{Molecular density functional theory calculations}

\subsection{Hydration free energy}

\begin{figure}[h]
\centering{}\includegraphics[width=8.5cm]{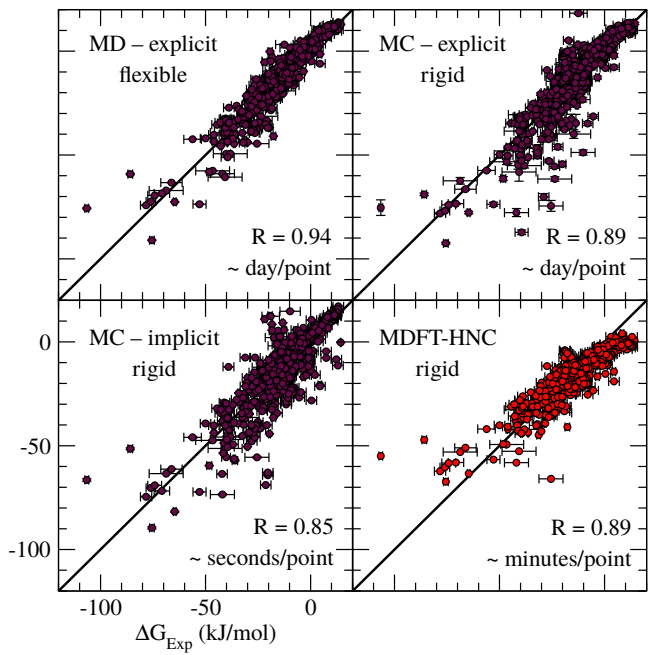}\caption{Correlations between experimental hydratation free energies and hydration
free energies obtained with (i) a flexible solute in explicit solvent
alchemical transformation simulations and with a rigid solute (ii)
in explicit solvent simulations, (iii) in implicit solvent simulations
and (iv) with MDFT-HNC calculations for the FreeSolv database.\label{fig:correlations}}
\end{figure}

Here we assess the quality of hydration free energies of small drug-like
molecules predicted by MDFT-HNC. In fig.~\ref{fig:correlations},
we show the correlation between the experimental HFE of the Freesolv
database as compared to those obtained by a MDFT-HNC calculations.
The results, that is, the output of the minimization of the functional
given in eq.~\ref{eq:minimisation}, were obtained within a cubic
supercell of side 32~Å, with periodic boundary conditions, a spacial
resolution of 0.25~Å (= 128 grid nodes in each direction) and an
angular resolution of 84 orientations per spatial grid node. The MDFT
calculations were performed on the initial configuration given in
the FreeSolv database. MDFT calculations did not converge for 36 molecules
(5 \% of the database, see SI for a list of these molecules) that
have a local charge too high for the HNC approximation (see SI for
more information). All results presented below are for the 606 molecules
that led to convergence. 

In fig.~\ref{fig:correlations}, we show the correlations between
experimental HFE, reference calculations in explicit and implicit
solvents, and those predicted by MDFT-HNC. Table~\ref{tab:corr_summary}
concatenates the statistical measures characterizing those four correlations.

\begin{table}[h]
\centering{}%
\begin{tabular}{c>{\centering}m{1.5cm}>{\centering}m{1.5cm}>{\centering}m{1.5cm}>{\centering}m{1.5cm}}
\toprule 
 & \centering{}Flexible solute$^{(a)}$ & \multicolumn{3}{c}{Rigid solute}\tabularnewline
\midrule 
 & \centering{}Explicit solvent & \centering{}Explicit solvent & \centering{}Implicit solvent & \centering{}MDFT\\
HNC\tabularnewline
\midrule 
MUE~(kJ/mol) & \centering{}$4.3$ & \centering{}$5.4$ & \centering{}$6.8$ & \centering{}$5.9$\tabularnewline
RMSE~(kJ/mol) & \centering{}$5.8$ & \centering{}$8.3$ & \centering{}$9.5$ & \centering{}$7.5$\tabularnewline
P-bias (\%) & \centering{}$-11.5$ & \centering{}$-3.2$ & \centering{}$-5.9$ & \centering{}$15.0$\tabularnewline
Pearson's $R$ & \centering{}$0.94$ & \centering{}$0.89$ & \centering{}$0.85$ & \centering{}$0.89$\tabularnewline
Spearman's $\rho$ & \centering{}$0.94$ & \centering{}$0.90$ & \centering{}$0.83$ & \centering{}$0.88$\tabularnewline
Kendall's $\tau$ & \centering{}$0.80$ & \centering{}$0.75$ & \centering{}$0.67$ & \centering{}$0.70$\tabularnewline
cpu-time~(s) & \centering{}$\sim10^{5}$ & \centering{}$\sim10^{5}$ & \centering{}$\sim10^{0}$ & \centering{}$\sim10^{2}$~$^{\textit{(b)}}$\tabularnewline
\bottomrule
\end{tabular}\caption{Summary of statistical measures characterizing the correlation between
the experimental hydration free energies and the ones obtained from
(\emph{i}) flexible solute $\equiv$ a mixture of conformers in explicit
solvent molecular dynamics simulations, and (\emph{ii}) from rigid
solute $\equiv$ single conformer calculation in reference explicit
solvent Monte Carlo simulations, or \textit{(iii)} implicit solvent
MC simulations and (\emph{iv}) MDFT-HNC, for the whole FreeSolv database.
Cross-correlations between computational methods are given in SI.
Statistical measures : MUE - mean unsigned error, RMSE - root-mean-squared
error, P-bias - percent bias, Pearson's correlation coefficient $R$
(measure of linear correlation), Spearman's and Kendall's ranking
correlation coefficients $\rho$ and $\tau$ (measures of monotonic
correlation). Orders of magnitude of cpu-times are given per solute
molecule. \textit{\emph{One can do thousands of MDFT or implicit solvent
calculations, not contrary to explicit solvent calculations. }}\textit{(a)
}Simulations done by Mobley \textit{et al. }\cite{mobley_2013}\textit{
(b)} Distribution of computing time is given in SI. \label{tab:corr_summary}}
\end{table}

MDFT-HNC has a mean unsigned error (MUE) to experimental values of
5.9~kJ/mol, while reference simulations MUE equals 5.4~kJ/mol. HFEs
are in all cases well (linearly and monotonically) correlated as can
be seen in fig.~\ref{fig:correlations} and confirmed with Pearson's
$R$ and Spearman's $\rho$ coefficients close to 0.9. The computational
cost, or average computation time, for each MDFT SFE lies within 5
minutes on a single cpu thread. The current implementation of MDFT
uses shared memory parallelism to decrease this time to less than
1 minute on any 2018's laptop that has 4 to 8 available threads. This
is 3 orders of magnitude faster than reference calculations, for a
MUE increased by 0.5 kJ/mol.

While taking few tens of seconds at most to predict SFE, parametric
implicit solvent simulations have a MUE of 6.8 kJ/mol, i.e. an increase
of 0.9 kJ/mol when compared with MDFT-HNC, and correlation constants
below 0.9. These results agree well with the deep review by Knight
and Brooks \cite{knight_2011}. Furthermore, it should be noted that
MDFT also gives the full molecular equilibrium solvent structure around
the solute, as shown in the next section, that can not be obtained
with implicit solvent MD simulations where the density is homogeneous
everywhere outside the solute core.

In summary, MDFT-HNC trades 0.12 kcal/mol of precision compared to
reference fully atomic simulations for a speed up of 3 orders of magnitude
at least. It takes less than an hour to compute HFEs for the whole
Freesolv database by MDFT-HNC. It is quick enough to be used in virtual
screening of millions of molecules which is not the case of explicit
solvent simulations and a large potential improvement with respect
to the methods, implicit solvent and scoring functions, used in drug
discovery. At the moment the virtual screening is a multistage process
with an initial screening stages with scoring functions to select
$\sim100$ hits and a final screening with MD simulations to obtain
few final leads. This multistage process could be replaced with a
unique MDFT-based stage.

\subsection{Solvent equilibrium structure, polarization and beyond}

\begin{figure}[h]
\begin{tabular}{ll}
(a) & (b)\tabularnewline
\includegraphics[height=4.2cm]{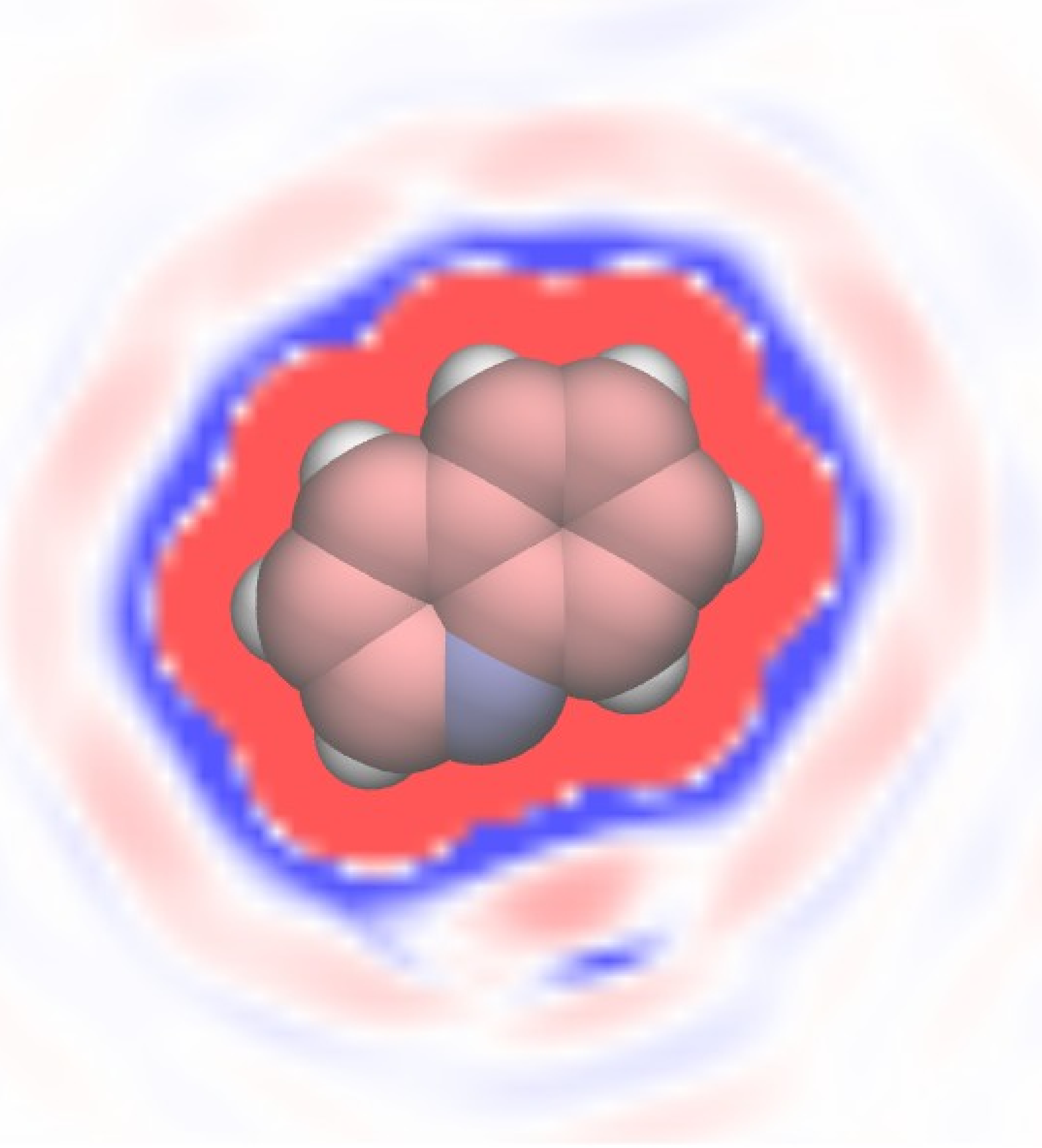} & \includegraphics[height=4.2cm]{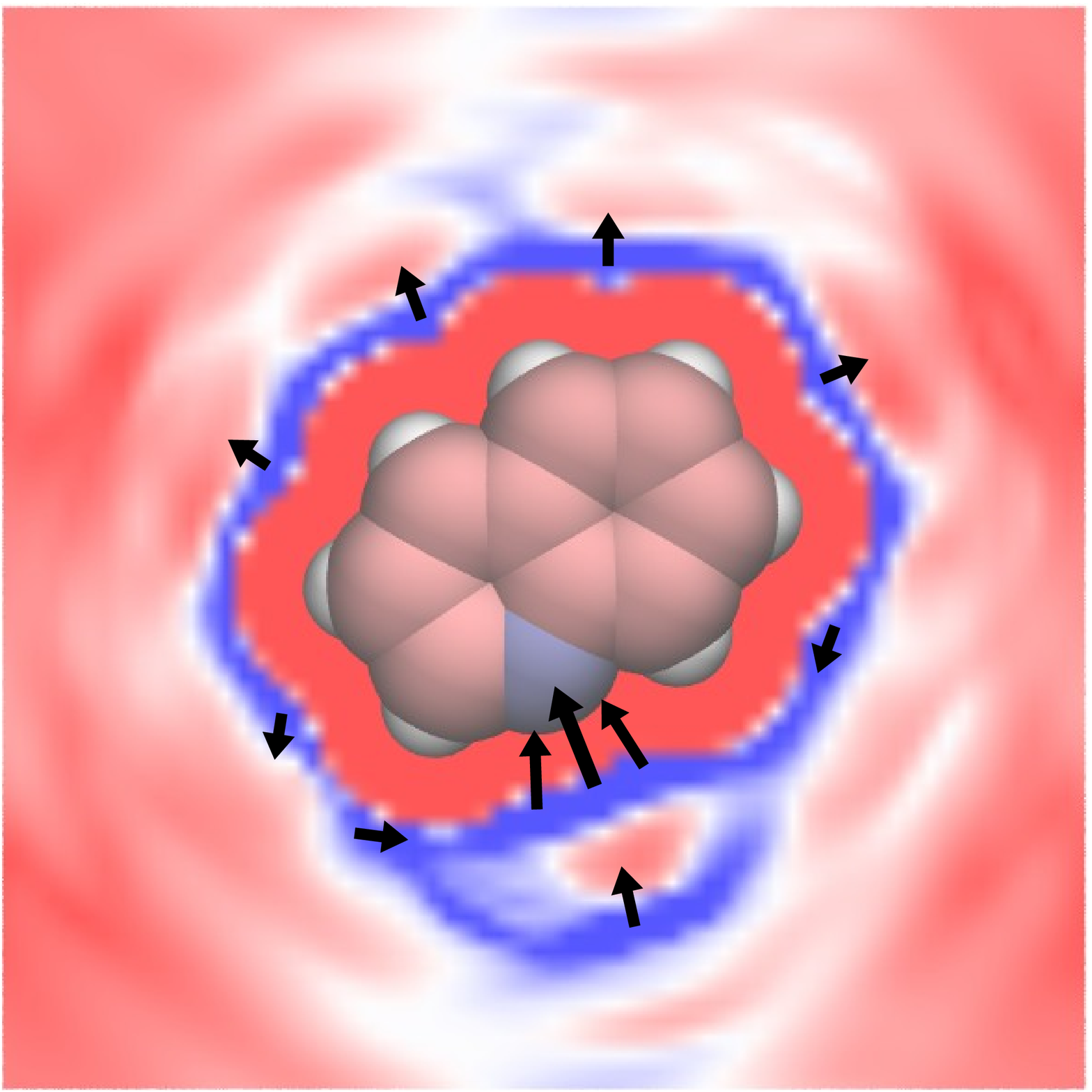}\tabularnewline
\end{tabular}\caption{(a) Water density map (red : $\rho<\rho_{\textrm{bulk}}$, white :
$\rho=\rho_{\textrm{bulk}}$ and blue : $\rho>\rho_{\textrm{bulk}}$)
and (b) norm of the polarization vector field (blue : high polarization,
black arrows representing the orientation) in the plane of the quinoline
molecule obtained obtained with MDFT-HNC \label{fig:maps}}
\end{figure}

\begin{figure}[h]
\includegraphics[width=8cm]{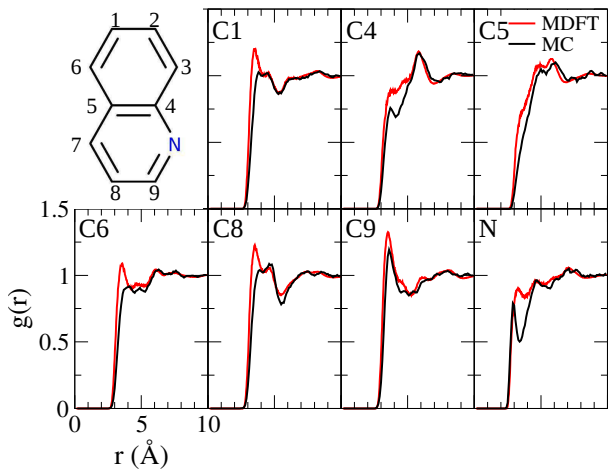}\caption{Chemical structure of the quinoline and the radial distribution function
$g(r)$ between the heavy atoms of the quinoline and water oxygens
(red lines correspond to MDFT-HNC as obtained in few minutes, and
black ones to MD simulations as obtained in few hours). \label{fig:rdf}}
\end{figure}

Equation~\ref{eq:minimisation} states that at the same time as MDFT
produces the solvation free energy of arbitrarily complex molecule
(the value of the functional at its minimum), it produces the equilibrium
solvent structure around this solute (the density that minimizes the
functional), in its full molecular description, $g(\boldsymbol{r},\omega)=\frac{\rho(\boldsymbol{r},\omega)}{\rho_{\textrm{bulk}}}$.
From this full molecular distribution, one can extract more readable
information. For instance, the first moment of $g(\boldsymbol{r},\omega)$
is the three dimensional scalar field $g(\boldsymbol{r})\equiv\frac{1}{8\pi^{2}}\int g(\boldsymbol{r},\omega)\textrm{d}\omega$
from which one can derive the usual spherically symmetric radial distribution
function $g_{i}(r)$ between solute sites and water oxygens. Fig.
\ref{fig:maps}(a) shows the three dimensional scalar field $g(\boldsymbol{r})$
in the plane of the quinoline molecule and fig. \ref{fig:rdf} shows
the radial distribution functions between the heavy atoms of quinoline
and solvent oxygen atoms. Fig.~\ref{fig:rdf} also shows the radial
distribution functions obtained from a explicit solvent simulation.

We see the solvation layer structure in fig. \ref{fig:maps}(a) and
radial distribution functions, fig. \ref{fig:rdf} given by the MDFT-HNC
corresponds well with the ones obtained with simulation : The first
peak of the rdfs are at the right distance, meaning that the first
solvation shell lies where it should. We observe a classical overestimation
with HNC of the height of the first peak. The $g(r)$ for nitrogen-oxygen
is qualitative only: The nitrogen atom wears a high partial charge
of $-0.65e$, which is difficult a case for HNC that does not perform
as well on strong charges. The typical overestimation of the height
of the first peak in the HNC approximation is due to the theory being
a second order density expansion of the functional, missing higher
order repulsion terms (packing effects) between excess densities \cite{levesque12_1}.

\begin{figure}[h]
\includegraphics[height=3cm]{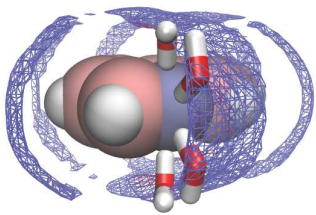}\caption{Representation of the quinoline molecule with four most probable water
molecules and the isosurface corresponding to $\rho=3\rho_{\textrm{bulk}}$.
\label{fig:quinoline}}
\end{figure}

Another important quantity embedded in $g(\boldsymbol{r},\omega)$
is the polarization field $\boldsymbol{P}(\boldsymbol{r})\equiv\int\hat{\omega}g(\boldsymbol{r},\omega)\textrm{d}\omega$
where $\hat{\omega}$ is the unitary vector along the dipole axis
depending on $(\theta,\phi)$ only. Fig. \ref{fig:maps}(b) shows
the $l_{2}$-norm of the polarization field, i.e., $\left\Vert \boldsymbol{P}(\boldsymbol{r})\right\Vert $,
in the plane of the quinoline molecule obtained with MDFT-HNC. As
expected we find high polarization close to the sites wearing localized
charges, and the expected polarization with OH pointing toward N.
One can also obtain so called water maps from $g(\boldsymbol{r},\omega)$,
catching the most probable water molecules position and orientation
around the solute. Fig. \ref{fig:quinoline} shows the four most probable
position and the orientation of water molecules around the quinoline
molecules. They are found close to the nitrogen atom. 

It should be noted that the equilibrium molecular solvent density
$g(\boldsymbol{r},\omega)$ is a direct output of MDFT. In the case
of molecular simulations, one would have to accumulate such data during
a long trajectory, averaging in a three-dimensional pixel (a voxel),
typically of width 0.5~Å. This can be tackled nowadays, especially
with the recent approach using also the forces to decrease the variance
of the estimate of $g$~\cite{borgis_2013,delasheras_2018}. Nevertheless,
accumulating data for a given orientation in the full six-dimensional
orientation and position space is a daunting task, even more difficult
than computing SFE. MDFT produces this 6-dimensional map in the same
few minutes as it needs to predict the SFE. 

\section{Conclusion }

The hydration or binding free energy of a drug-like molecule is a
key data for early stage drug discovery. This quantity is nowadays
calculated either (i) by fast but inaccurate scoring functions or
implicit solvent methods, or (ii) by accurate but slow atomistic simulations
and free energy methods. In this paper, we have shown on the 600+
drug-like molecules of the FreeSolv database that the molecular density
functional theory (MDFT) in its simplest hypernetted-chain (HNC) approximation
is able to trade 0.5 kJ/mol (0.1~kcal/mol) of mean unsigned error
compared to the most precise simulation for a speed up of five orders
of magnitude for the solvation free energies. MDFT can thus be used
directly during the screening and docking process. Furthermore, since
MDFT-HNC is exact up to second order in density correlations, it can
only be improved by bridge functionals. Last but not least, the MDFT
also predicts the full molecular, angular, three-dimensional solvent
density, also called water maps, around the solute or protein-ligand
complexes, spotting where are the most important water molecules.

\bibliographystyle{unsrt}
\bibliography{biblio}

\end{document}